\documentclass[aps,pre,twocolumn,floatfix]{revtex4}
\usepackage[pdftex]{color}
\usepackage{latexsym,hyperref,amssymb,amsmath,gensymb}
\usepackage{graphicx,dcolumn,bm}
\def\geqsim{\lower.73ex\hbox{$\sim$}\llap{\raise.4ex\hbox{$>$}}$\,$}
\def\leqsim{\lower.73ex\hbox{$\sim$}\llap{\raise.4ex\hbox{$<$}}$\,$}

\newcommand{\thalf}{{\textstyle{\frac{1}{2}}}}

\newcommand{\bos}{\boldsymbol}
\newcommand{\tbf}{\textbf}
\newcommand{\tit}{\textit}
\newcommand{\tsf}{\textsf}

\newcommand{\mbf}{\mathbf}
\newcommand{\msf}{\mathsf}

\newcommand{\beq}{\begin{equation}}
\newcommand{\eeq}{\end{equation}}
\newcommand{\bea}{\begin{eqnarray}}
\newcommand{\eea}{\end{eqnarray}}
\newcommand{\barr}{\begin{array}}
\newcommand{\earr}{\end{array}}
\newcommand{\bean}{\begin{eqnarray*}}
\newcommand{\eean}{\end{eqnarray*}}
\newcommand{\bei}{\begin{itemize}}
\newcommand{\eei}{\end{itemize}}
\newcommand{\ben}{\begin{enumeration}}
\newcommand{\een}{\end{enumeration}}
\newcommand{\bec}{\begin{center}}
\newcommand{\eec}{\end{center}}
\newcommand{\nn}{\nonumber}
\newcommand{\tsc}{\textsc}

\newcommand{\lt}{\left}
\newcommand{\rt}{\right}
\newcommand{\ep}{\epsilon}
\definecolor{navyblue}{rgb}{.05,0,.55}
\newcommand{\tcr}[1]{\textcolor{red}{#1}}

\newcommand{\tcg}[1]{\textcolor{green}{#1}}

\begin{document}
\begin{widetext}
\begin{flushleft}
{\Large{\textbf{\textsf{Cell-Penetrating Peptides, Electroporation, and Drug Delivery}}}}
\end{flushleft}

\begin{flushleft}
\textsf{Kevin Cahill\hfill\today\\cahill@unm.edu}\\
\textsf{Biophysics Group,
Department of Physics \& Astronomy,
University of New Mexico, Albuquerque, NM 87131}
\end{flushleft}

\pagestyle{myheadings}
\markright{Cell-Penetrating Peptides and Drug Delivery}

\begin{abstract}\noindent
\textsf{ABSTRACT \quad 
Certain short polycations, 
such as TAT and oligoarginine, 
rapidly pass through the plasma membranes
of mammalian cells 
by a mechanism called transduction,
as well as by endocytosis and macropinocytosis.
These cell-penetrating peptides (CPPs)
can carry with them cargos of 
30 amino acids, more than 
the nominal limit of 500 Da and enough
to be therapeutic.
An analysis of the electrostatics
of a charge outside the cell membrane
and some recent experiments
suggest that transduction may proceed
by molecular electroporation.
Ways to target diseased cells,
rather than all cells, are discussed.}
\end{abstract}

\maketitle
\end{widetext}
\section{The Problem of Drug Delivery
\label{The Problem of Drug Delivery}}

We could cure cancer
if we knew how to deliver a drug
intact to the cytosol 
of every cancer cell,
sparing healthy cells.
The circulatory system can deliver
a drug to every cell in the body,
and certain chemical tricks 
protect drugs from
peptidases~\citep{Goodman1993} 
and nucleases~\citep{Rajeev2005}\@.
But it's harder to cope
with antibodies, spare
healthy tissues, and get
drugs past the plasma membrane,
which blocks or endocytoses  
molecules in excess of 
500 Da~\citep{Lipinski1997}\@.
\par
This paper is about 
cell-penetrating peptides and
other cations that can 
overcome the 500-Da restriction barrier 
and about tricks that may spare
healthy cells.
Section~\ref{Cell-Penetrating Peptides}
describes several 
cell-penetrating peptides (CPPs),
and section~\ref{Therapeutic Applications} 
sketches a variety of therapeutic applications
of CPPs.  Section~\ref{Mammalian Plasma Membranes}
reviews basic facts about the lipid
bilayer of the eukaryotic cell.
New work on the electrostatics
of the bilayer is presented in 
section~\ref{Membrane Electrostatics}\@.
Section~\ref{Electroporation Model} sketches
a model~\citep{Cahill2009,Cahill2010} 
of the transduction of polyarginine and
mentions some experimental
support~\citep{Herce2009}
which the model has recently received.
Section~\ref{Smart Drugs} discusses
a broader class of cell-penetrating molecules
and suggests ways to target cancer cells.

\section{Cell-Penetrating Peptides
\label{Cell-Penetrating Peptides}}
In 1988, two groups~\citep{Loewenstein1988,Pabo1988} 
working on HIV
reported that the \tcr{t}rans-\tcr{a}ctivating 
\tcr{t}ranscriptional activator
(\tcr{TAT}) of HIV-1 can cross cell membranes.
The engine driving this 86-aa 
cell-penetrating peptide (CPP)
is residues 48--57 
\textsc{grkkrrqrrr}
which carry a charge of +8\(e\)\@.
Other CPPs were soon found.
Antp (aka Penetratin, PEN) is residues 43--58
\(\textsc{rqikiwfqnrrmkwkk}\)
of Antennapedia, a homeodomain
of the fly; it carries a charge of +7\(e\)\@.
The polyarginine R\(^n\) carries charge
\(+ n e \), where often \(n = 7, 8\), or 9\@.
Other CPPs have been discovered
(VP22) or synthesized (transportan)\@.
The structural protein VP22
of the tegument of herpes simplex virus type 1 (HSV-1)
has charge +15\(e\)\@.
Transportan
\(\textsc{gwtln}\textsc{sagyllg}\)-\(\textsc{k}\)-\(\textsc{inlkala}\textsc{alakk}\textsc{il}\)-amide
is a chimeric peptide constructed from 
the 12 N-terminal residues of galanin 
in the N-terminus with the 14-residue
sequence of mastoparan  
and a connecting lysine~\citep{Lindberg2001}\@.
With its terminal amide group,
its charge is +5\(e\)\@.
\par
These and other short, positively
charged peptides can penetrate 
the plasma membranes of live cells
and can tow along with them cargoes
that greatly exceed the 500 Da 
restriction barrier.
They are promising therapeutic 
tools when towing cleverly chosen peptide cargoes 
of from 8 to 33 amino acids~\citep{Tsien2004,Dowdy2005,Pugh2002,Kaelin1999,Fong2003,Cohen2002,Robbins2001,Datta2001,Hosotani2002,Hsieh2006,SnyderDowdy2004,Morano2006,Tuennemann2007}\@.
\par
Many early experiments on CPPs 
were wrong because the
cells were fixed or insufficiently
washed.  Even careful experiments 
sometimes have yielded inconsistent 
results---in part
because fluorescence varies with the 
(sub)cellular conditions and 
the fluorophores~\citep{Seelig2007}.
\par
Yet some clarity is emerging:
TAT carries cargoes across 
cell membranes with high efficiency
by at least two functionally distinct mechanisms
according to whether the cargo is big
or small~\citep{Cardoso2006}\@.
Big cargoes, such as proteins
or quantum dots, enter via 
caveolae endocytosis
and macropinocytosis~\citep{Dowdy2004,Brock2007},
and relatively few escape the 
cytoplasmic vesicles in which they then
are trapped~\citep{Cardoso2006}\@.
\par
Small cargoes, such as peptides 
of fewer than 30--40 amino acids, 
enter both slowly
by endocytosis and rapidly by transduction
with direct access to the cytosol,
an unknown mechanism
that uses the membrane 
potential~\citep{Cardoso2006,Prochiantz2000,Rezsohazy2003,Wender2005,Shen2005}\@.
Peptides fused to TAT
enter cells
within seconds~\citep{Seelig2005}\@.
\par
It remains unclear how big cargoes
aided by several CPPs enter cells~\citep{Patel2007}\@.
For instance,
superparamagnetic nanoparticles
encased in aminated dextran and
attached to 45 tat peptides are
thought to enter cells by
adsorptive endocytosis\citep{Fawell1994,Nagahara1998,Bulte2006} 
but they do enter slowly at 
4\(^\circ\) C~\citep{Garden2006}\@.

\section{Therapeutic Applications
\label{Therapeutic Applications}}

The use of a cell-penetrating peptide (CPP)
to carry into cells a biologically active peptide 
of up to some 35 amino acids (aa) allows
for \(20^{35} = 3 \times 10^{45}\) different peptides and
may lead to strikingly smart drugs,
some of which may selectively target 
tumor cells~\citep{Tsien2004}\@.
I will now briefly describe 11 
promising advances toward this goal of
cell-penetrating-peptide drugs (CPPDs)\@.
\par
The CXC chemokine receptor 4 (CXCR4)
is overexpressed in \(>\!\!\,\!\,20\) types
of cancer, including prostate, breast,
colon, and small-cell lung cancer.
Snyder \textit{et al.}~\citep{Dowdy2005}
attached the CXCR4 ligand DV3 to the CPP TAT
and then either to
a Cdk2-antagonist peptide
(DV3-TAT-RxL)
or to
a p53-activating peptide
(DV3-\(\{\)TAT-p53C'\(\}_{ri}\)) 
in which \textit{ri} means 
retroinverso~\citep{Goodman1993}\@.
The DV3+RxL and DV3+p53C' cargoes 
respectively consisted of 19 and 33 aa.
Both ligand-guided CPPs were more than
twice as effective as unguided CPPs
in killing CXCR4 expressing Namalwa
lymphoma cells.
\par
The transcription factor hypoxia-inducible factor-1
(HIF), a master regulator of the hypoxic response,
is itself regulated thru the oxygen-dependent 
degradation domains (ODD) of 
its \(\alpha\)-chains (HIF\(\alpha\))\@.
The NODD and CODD peptides 
respectively are the amino- and carboxyl-terminal 
sequences of ODD\@. 
Pugh \textsl{et al.}
injected TAT fused to CODD (tat-CODD)
into sponges implanted subcutaneously (s.c.) in mice.
After 7 days, they found that tat-CODD
but not the control
mutant tat-CODD\(^{\mbox{mut}}\) produced
blood vessels of increasing density and 
complexity~\citep{Pugh2002}\@.
Their results for tat-NODD were similar.
Thus TAT delivery of NODD 
and CODD peptides
can stimulate angiogenesis 
and may lead to 
therapies for ischemic diseases.
The cargoes NODD and CODD respectively
were 28 and 19 aa long.
\par
If the DNA-binding ability of 
the transcription factor E2F1
is not curtailed as cells traverse 
and prepare to exit S 
phase, then apoptosis is likely.
In normal cells, the retinoblastoma 
tumor-suppressor protein pRB converts E2Fs from
activators to repressors of transcription.
But pRB often is not present in transformed cells.
Cyclin A/cdk2 also neutralizes E2F1's DNA-binding ability. 
So if one blocked the interaction of cyclin A/cdk2\
and E2F1 in cells,
then the transcription factor E2F1
would continue to activate transcription
after exit from S phase in cancer cells
but not in normal cells equipped,
as they are, with pRB\@.
Apoptosis then would occur in the transformed cells
but not in the normal ones.
Chen \textit{et al.}~\citep{Kaelin1999}
used TAT and PEN to
carry the synthetic peptides
\tsc{pvkrrlfg} and \tsc{pvkrrldl},
which block the interaction of cyclin A/cdk2
and E2F1, into
U2OS osteosarcoma cells,
T-antigen-transformed WI38/VA13 cells,
and healthy WI38 cells.
They found that apoptosis occurred
in and only in the cancer cells~\citep{Kaelin1999}\@.
Both cargoes were 8 aa long.
\par
The CPPD PEN-\tsc{pvkrrldl} injected
in and near tumors
in nude mice that had been s.c.\ injected with SVT2 cells
with altered cyclin D/Rb pathways produced large areas
of apoptosis and necrosis, particularly at the tumor
periphery~\citep{Fong2003}\@. 
PEN-\tsc{pvkrrlfg} had similar effects 
on herceptin-resistant mammary tumors 
from HER2 transgenic mice implanted
in syngeneic FVB mice~\citep{Fong2003}\@.
Both cargoes were 8 aa long. 
\par
HDM2 binds and inhibits p53,
keeping it at low levels in normal cells.
Oncogenic mutations disrupt this HDM2-p53
equilibrium, allowing p53 to accumulate and 
induce stasis or apoptosis.
In most cancer cells,
p53 is mutated
or HDM2 is overexpressed.
Uveal melanomas (common eye
cancers in adults)
overexpress HDM2\@.
A peptide that blocks HDM2 may liberate p53,
which may induce apoptosis in cancer cells
but remain inactive in normal ones.
The sequence \tsc{qetfsdlwkllp} (\(\alpha\)HDM2)
of the p53 binding site 
for HDM2 blocks HDM2\@.   
TAT-G-\(\alpha\)HDM2 at
concentrations of 200--300 \(\mu\)M 
induces apoptosis in MM-23, MM-24, \& MM-26 uveal,
Y79 \& WERI retinoblastoma, 
U2OS osteosarcoma, and
C33A cervical cancer cells
with little effect on normal cells~\citep{Cohen2002}\@.
Injection of \(10^7\) WERI human 
retinoblastoma cells produced intraocular tumors
in the anterior chambers of rabbit eyes.
But the injection of
TAT-\(\alpha\)HDM2 
(intraocular concentration 200 \(\mu\)M)
began to dissolve the tumors 
into a fine cloud within 24 hours.
A second injection reduced the viable 
tumor mass by 76\% within 72 hours 
with no histologic damage to
other ocular tissues.
The cargo was 13 aa long.
\par
The CPP \tsc{rrqrrtsklmkr} (``PTD-5,'' charge +7\(e\)) 
joined to the antimicrobial peptide 
\tsc{klaklakklaklak} (``KLA'')
with a diglycine spacer forms 
the pro-apoptotic peptide 
PTD-5-GG-KLA (``DP1'')\@.
Robbins \textit{et al.}\ injected 50 \(\mu\)L
of 1 m\textsc{m} DP1 or KLA for 11 days 
into C57BL/6 mice with single-flank,
day-12 MCA205 fibrosarcomas.
After only 8 days of treatment,
DP1 but not KLA had shrunk the tumors.
DP1 but not KLA or PTD-5
activated caspase-3 and mediated 
apoptosis~\citep{Robbins2001}\@. 
The cargo was 16 aa.
\par
Renal-cell cancer (RCC) is the seventh most common 
and is resistant to non-surgical therapies.
Sporadic clear-cell renal cancer often exhibits
functional inactivation of the protein pVHL
of the von-Hippel-Lindau (VHL) gene.
Residues 104--123 of a \(\beta\)-sheet of pVHL
inhibits the IGF-I signaling upon which
RCC is dependent~\citep{Datta2001}\@.
Daily intraperitioneal injections of 2 nmol
of TAT-FLAG-pVHL(104--123) arrested and then reduced
tumors of 786-O RCC cells in nude mice, 
while TAT-FLAG had no effect.
(FLAG is the antigen \tsc{ykddddk}\@.)
The cargo was 27 aa long.
\par
The tumor-suppressor gene 16INK4A
often is inactivated 
by intragenic mutation,
homozygous deletion, or 
methylation silencing in many human cancers,
especially pancreatic cancer.
Its protein p16 inhibits the 
phosphorylation of Rb by cdk-4 and by cdk-6,
and so p16 blocks the G\(_1\)\(\to\)S phase transition.
Residues 84--103 of p16 are sufficient to
block this transition.  
The Trojan p16 peptide was
p16(84--103)C linked by a disulfide bond
to C-Antp:
\tsc{daaregfldtlvvlhragarc-crqikiwfqnrrmkwkk}\@.
Trojan p16 (i.p.~100 \(\mu\)g/mouse/day)
reduced AsPC-1 and BxPC-3 
s.c.\ tumors respectively by
factors of 2 and 5 without hematological cytotoxicity
or body weight loss~\citep{Hosotani2002}\@. 
The cargo was 22 aa long.
\par
The DOC-2/DAB2 (differentially expressed in
ovarian cancer-2/disabled 2) protein
often is lost in prostate cancer.
As part of the homeostatic machinery
in the normal prostate epithelium,
DOC-2/DAB2 modulates the Grb2-SOS-MAPK
signal axis.  The small peptide
\tsc{fqlrqpplvpsrkge} is less immunogenic
than the protein DOC-2/DAB2 but still
has some of its ability to interact with SH3 domains.
Hsieh \textit{et al.}~used
the CPP R11 to carry this peptide into cells.
The CPPD \tsc{r\(^{11}\)-ggg-fqlrqpplvpsrkge}
at 5 \(\mu\)M for 3 hours
inhibited the growth of LNCaP and C4-2 
prostate-cancer cells.~\citep{Hsieh2006}
The cargo was 18 aa long.
\par
The peptide p53C\(^\prime\), derived from the K-rich 
C-terminal domain of p53, 
activates specific DNA binding by p53,
activates wild-type p53, 
restores function to several p53 contact mutants, and
induces apoptosis or G\(_1\) growth arrest 
in cancer cells
but not in normal ones~\citep{SnyderDowdy2004}\@.
The retro-inverso~\citep{Goodman1993} 
D-isomer TAT-p53C\(^\prime_{ri}\)
\tcr{rrrqrrkkrgy}\tcg{gkkhrstsqgkksklhssharsg}
resists proteolysis
better than the L-isomer 
p53C\(^\prime\)-TAT\@.
TAT-p53C\(^\prime_{ri}\)
induces G\(_1\) cell-cycle arrest and senescence 
better than p53C\(^\prime\)-TAT
in murine TA3/St mammary cancer cells,
which express wild-type p53, and
induces the transcription of p53 from a wild-type
p53 expression vector 
in p53-null, H1299 lung-cancer 
cells~\citep{SnyderDowdy2004}\@.
Injections (i.p.) of 600 \(\mu\)g of TAT-p53C\(^\prime_{ri}\)
once a day for 12 days inhibited 
solid-tumor growth in mice and
led to a 6-fold increase 
in longevity in mice with
terminal peritoneal carcinomatosis.
Sixteen injections (i.p.) of 900 \(\mu\)g 
of TAT-p53C\(^\prime_{ri}\)
over 20 days cured 50\% 
of mice with terminal 
peritoneal lymphoma~\citep{SnyderDowdy2004}\@.
The cargo was 23 aa long.
\par
Residues 1--15 of the 
human ventricular myosin light chain-1 (VLC-1)
binds to actin, targets the 
actin/MLC-1 interaction, and
improves the contraction of
isolated perfused hearts~\citep{Morano2006}\@.
VLC-1 fused to TAT entered 
adult cardiomyocytes with high efficiency,
accumulated
in the actin-containing I-band
of their sarcomeres,
and enhanced their contractility
without changing their myoplasmic
Ca\(^{2+}\) levels~\citep{Tuennemann2007}\@.
The cargo was 15 aa long.
\par
In these transduction experiments,
the heaviest cargo was 33 amino acids.
In my somewhat casual literature search,
I found no biomedical articles describing
the transduction of heavier cargoes.  
In the biophysical experiments,
the cargoes respectively were 20, 22, and 26 aa 
(apart from a tiny rhodamine tag)~\citep{Cardoso2006},
and just a \(\sim\!\,\)400 Da 
fluorophore~\citep{Cardoso2008}.

\section{Mammalian Plasma Membranes
\label{Mammalian Plasma Membranes}}
The plasma membrane of a mammalian cell
is a lipid bilayer that is 4 or 5 nm thick.
Of the four main phospholipids in it,
three---phosphatidylethanolamine (PE),
phosphatidylcholine (PC), and
sphingomyelin (SM)---are neutral,
and one, phosphatidylserine (PS), 
is negatively charged\@.
In live cells, PE and PS 
are mostly in the cytosolic layer,
and PC and SM in the 
outer layer~\citep{Zwaal1999,MBoC4587}\@.  
Aminophospholipid translocase (flippase) moves 
PE and PS 
to the inner layer;
floppase slowly moves all phospholipids
to the outer layer~\citep{Zwaal1999}\@.
\par
Glycolipids make up about 5\%
of the lipid molecules of the outer layer
of a mammalian plasma membrane
where they may form lipid rafts.
Their hydrocarbon tails normally
are saturated.  Instead of a modified
phosphate group, they are decorated 
with galactose, glucose, 
GalNAc = N-acetylgalactosamine,
and other sugars.  The most complex
glycolipids---the gangliosides---have 
negatively charged 
sialic-acid (NANA) groups\@.
\par
A living cell maintains an
electrostatic potential of 
between 20 and 120 mV
across its plasma membrane.
The electric field \(E\) within
the membrane points into the cell
and is huge, about 15 mV/nm or
\(1.5 \times 10^7\) V/m
if the potential difference 
is 60 mV across a membrane of 4 nm.
Conventionally, one reports membrane
potentials as the electric potential
inside the cell minus that outside,
so that here \(\Delta V = - 60\) mV\@.
Near but outside the membrane,
this electric field falls-off
exponentially \(E(r) = E \, \exp(-r/D_\ell)\)
with the ratio of the distance \(r\)
from the membrane to the Debye length \(D_\ell\),
which is of the order of a nanometer\@.
The rapid entry of TAT fused to peptides is frustrated
only by agents that destroy the
electric field \(E\)~\citep{Cardoso2006}\@.
\par
Most of the phospholipids 
of the outer leaflet of the plasma membrane
are neutral PCs \& SMs\@.
They vastly outnumber
the negatively charged gangliosides, 
which are a subset
of the glycolipids, which themselves amount only
to 5\% of the outer layer.
Imagine now that 
polyarginine-cargo molecules
are in the extra-cellular environment.
Many of them 
will be pinned down by the electric field
\(E(r)\) just outside the membrane,
their positively charged guanidinium groups
interacting with the negative phosphate
groups of neutral
dipolar PC \& SM head groups~\citep{Wender2005}.
(Other CPP-cargo molecules 
will stick to negatively
charged gangliosides and  
to glycosaminoglycans (GAGs) 
attached to transmembrane proteoglycans (PGs);
these slowly will be endocytosed.
PGs with heparan-sulfate GAGs are needed
for TAT-protein endocytosis~\citep{Giacca2001}\@.)
It is crucial that the 
dipolar PC \& SM head groups 
are neutral and so do not
cancel or reduce the positive electric charge
of a CPP-cargo molecule.
The net positive charge of a CPP-cargo
molecule and the negatively charged
PSs under it on the inner leaflet
form a kind of capacitor.

\section{Membrane Electrostatics
\label{Membrane Electrostatics}}

A recent calculation~\citep{Cahill2010} 
of the electrostatic
potential due to a charge outside 
the phospholipid bilayer of a eukaryotic cell
shows how effectively this plasma membrane 
insulates the cell from external charges.
\par
The electrostatic potential 
in the lipid bilayer \(V_\ell(\rho,z)\)
due to a charge \(q\) at the point
\((0,0,h)\) on the \(z\)-axis, a
height \( h \) above
the interface between the lipid bilayer 
and the extra-cellular environment, is
\bea
V_\ell(\rho,z) & = &
\frac{q}{4\pi\epsilon_0 \epsilon_{w\ell}}
\, \sum_{n=0}^\infty
(p p')^n \lt(
\frac{1}{\sqrt{\rho^2 + (2nt+h-z)^2}} \rt. \nn\\
& & \lt. \mbox{} 
- \frac{p'}{\sqrt{\rho^2 + (2(n+1)t+h+z)^2}}\rt)
\label {Vellt}
\eea
in which \(t\) is the thickness
of the lipid bilayer,
\( \epsilon_{w\ell}
= ( \epsilon_w + \epsilon_\ell )/2 \)
is average of the relative permittivities
of the extra-cellular fluid \( \epsilon_w \)
and the lipid bilayer \( \epsilon_\ell \),
and \(p\) and \(p'\) are the ratios
\beq
p = \frac{\epsilon_w - \epsilon_\ell}
{\epsilon_w + \epsilon_\ell}
\quad \mbox{and} \quad
p' = \frac{\epsilon_c - \epsilon_\ell}
{\epsilon_c + \epsilon_\ell}
\label {p and p' t}
\eeq
which lie between 0 and 1,
\(\epsilon_c\) being the relative
permittivity of the cytosol~\citep{Cahill2010}\@.
The potential in the extra-cellular medium is
\bea
V_w(\rho,z) & = & 
\frac{q}{4\pi\epsilon_0 \epsilon_w}
\lt(\frac{1}{r}
+ \frac{p}{\sqrt{\rho^2 + (z+h)^2}} 
\rt. \nn\\
& & \mbox{} -  \lt. 
\frac{\ep_w \ep_\ell}{\ep_{w\ell}^2}
\sum_{n=1}^\infty
\frac{p^{n-1} p^{\prime n}}
{\sqrt{\rho^2 + (z + 2nt + h)^2}} \rt)
\label {Vex}
\eea
in which \(r =\sqrt{\rho^2 + (z-h)^2}\) 
is the distance from the charge \(q\)~\citep{Cahill2010}\@.
The potential in the cytosol 
due to the same charge \(q\) is
\beq
V_c(\rho,z) = \frac{q \, \ep_\ell}
{4\pi\ep_0\ep_{w\ell}\ep_{\ell c}}
\! \sum_{n=0}^\infty
\frac{(p p')^n}{\sqrt{\rho^2 + (2nt+h-z)^2}}.
\label {Vct}
\eeq
where \(\ep_{\ell c}\) is
the mean relative permittivity
\(\ep_{\ell c} = ( \ep_\ell + \ep_c )/2\)~\citep{Cahill2010}\@.
\par
If one seeks the potential only
directly above or below the charge,
that is, for \(\rho = 0\), then
these formulas become simpler 
and may be expressed in terms
of the Lerch transcendent
\beq
\Phi(z,s,\alpha) = \sum_{n=0}^\infty
\frac{z^n}{(n + \alpha )^s}.
\label {Lerch trancendent}
\eeq
For \(\rho = 0\),
the potential is 
\bea
V_\ell(0,z) & = & \frac{q}
{4\pi\epsilon_0 \epsilon_{w\ell}}
\, \sum_{n=0}^\infty
(p p')^n \lt(
\frac{1}{2nt+h-z} \rt. \nn\\
& & \lt. \mbox{} 
- \frac{p'}{2(n+1)t+h+z}\rt)
\nn\\
& = & \frac{q}{8\pi\epsilon_0 \epsilon_{w\ell} t}
\,  \lt[ \Phi\lt(pp',1,\frac{h-z}{2t}\rt) \rt.
\nn\\
& & \lt. \mbox{} - p' 
\Phi\lt(pp',1,1 + \frac{h+z}{2t}\rt) \rt]
\label {Vellt0}
\eea
in the lipid bilayer,
\bea
V_w(0,z) & = & \frac{q}{4\pi\epsilon_0 \epsilon_w}
\lt[ \frac{1}{|z-h|} + \frac{p}{z+h} \rt. 
\label {Vex0}\\
& & \lt. \mbox{} 
- \frac{\ep_w \ep_\ell}{\ep_{w\ell}^2 p}
\lt( \frac{1}{2t} \Phi(pp',1,\frac{z+h}{2t})
- \frac{1}{z+h} \rt) \rt]
\nn
\eea
in the extracellular environment, and
\beq
V_c(0,z) = \frac{q \, \ep_\ell}
{8\pi\ep_0\ep_{w\ell}\ep_{\ell c} t} 
\Phi(pp',1,\frac{h-z}{2t}).
\label {Vct0}
\eeq
in the cytosol.
\par
In and near the extracellular region,
these potentials are fairly well
approximated by the simple formulas 
\bea
V_w(0,z) & \approx &
\frac{q}{4\pi\epsilon_0 \epsilon_w}
\lt(\frac{1}{|z-h|}
+ \frac{p}{z+h} \rt)
\label {Vw simple}\\
V_\ell(0,z) & \approx &
\frac{q}{4\pi \epsilon_0 \epsilon_{wl} \, |z-h|} 
\label {simple formula}
\eea
which hold when the lipid bilayer
is infinitely thick.  
But the potential \(V_\ell(0,z)\)
drops significantly below the simple formula
(\ref{simple formula})
as \(z\) descends deeper into the
bilayer and nearly vanishes
at the lipid-cytosol
interface as does \(V_c(0,z)\) 
in the cytosol.

\begin{figure}
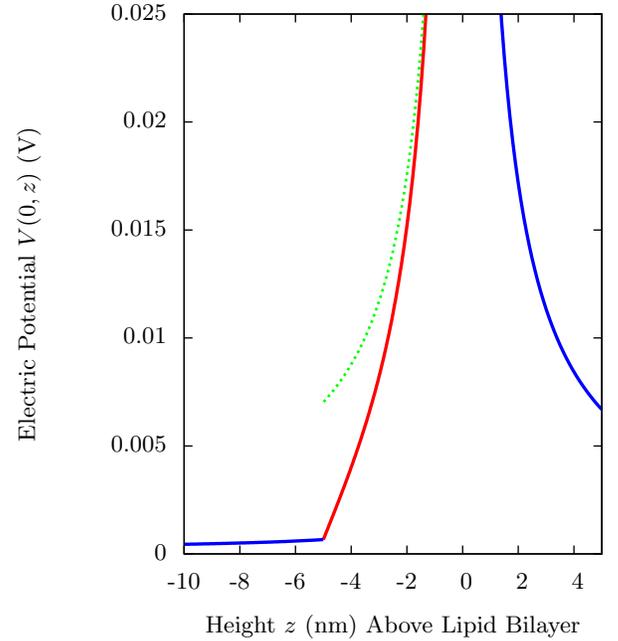

\centering
\input Vz
\caption{The potentials 
(\ref{Vellt0}--\ref{Vct0}) above, 
\(V_w(0,z)\) (blue), and below,
\(V_\ell(0,z)\) (red) and \(V_c(0,z)\) (blue),
a unit charge \(q=|e|\) 
at \((\rho,z) = (0,0)\)
are plotted for a lipid bilayer
of thickness \(t = 5\) nm\@.
The green dotted curve follows the
simple approximation (\ref{simple formula})\@.}
\label {Vzfig}
\end{figure}

\begin{figure}
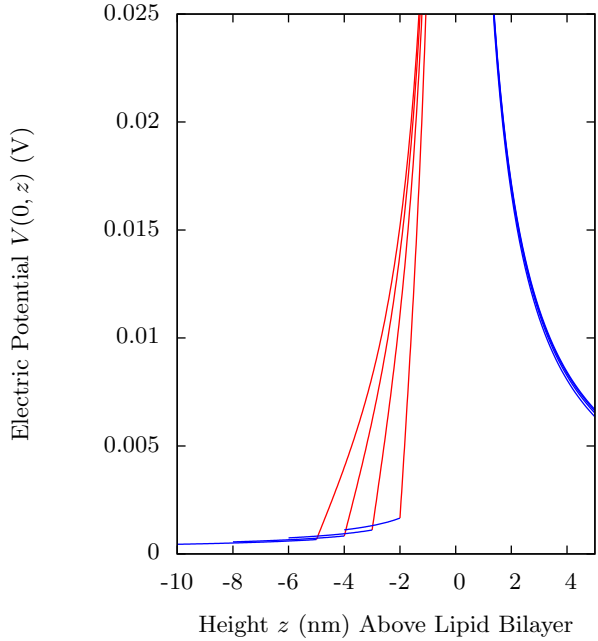

\centering
\input kT
\caption{The potentials 
(\ref{Vex0}, \ref{Vellt0},
\& \ref{Vct0}) above 
\(V_w(0,z)\) (blue) and below
\(V_\ell(0,z)\) (red) and \(V_c(0,z)\) (blue)
a unit charge \(q=|e|\) 
at \((\rho,z) = (0,0)\)
are plotted for lipid bilayers
of thickness \(t=2\), 3, 4, and 5 nm\@.
Note that the potential at the
interface between the cytosol
and the bilayer is less than
a tenth of \(kT\) even for
a bilayer as thin as 2 nm\@.}
\label {kTfig}
\end{figure}

\par
For a unit
charge \(q=|e|\) 
at \((\rho,z) = (0,0)\), 
the potentials (\ref{Vellt0}--\ref{Vct0}) 
are plotted
in Fig.~\ref{Vzfig}:
\(V_w(0,z)\) (blue), 
\(V_\ell(0,z)\) (red), and \(V_c(0,z)\) (blue)\@.
The green dotted curve follows the
simple approximation (\ref{simple formula})
to \(V_\ell(0,z)\)\@.
The potential \(V_\ell(0,-t)\)
at the interface between the cytosol
and the bilayer
is \(6.6 \times 10^{-4}\)~V,
so that a unit charge there would
have an electrostatic energy 
(due to the charge \(|e|\) at \((0,0)\))
of only about \(kT_b/40\) at body temperature.
\par
Fig.~\ref{kTfig} plots the potentials 
(\ref{Vellt0}--\ref{Vct0}) due to
a unit charge 
at \((\rho,z) = (0,0)\)
for lipid bilayers
of thicknesses \(t = 2\), 3, 4, and 5 nm\@.
Even for a bilayer as thin as 2 nm,
the potential \(V_c(0,-t)\) directly below
the charge at the interface between the cytosol
and the bilayer is less than
a tenth of \(kT_b/|e|\)\@.
\par
These potentials and figures
illustrate the extraordinary
electrostatic insulation
provided by the lipid bilayer.

\section{Electroporation Model
\label{Electroporation Model}}

\subsection{Electroporation
\label{Electroporation}}

Somewhat paradoxically,
the remarkable electrostatic insulation
provided by the lipid bilayer
makes membrane proteins
more sensitive to charges lying 
outside the cell and also
makes cells more 
vulnerable to electroporation.
The reason for both of these effects
is that the electric field \(\bos{E}\)
in the lipid bilayer due to a 
charge \(q\) at the interface between 
the bilayer and the extracellular
environment is 
\beq
\bos{E} = - \lt[ V_\ell(0,0) - V_\ell(0,-t) \rt]
\, \frac{\bos{\hat z}}{t}
\label {E}
\eeq
apart from the screening effects
of counterions.  Since even a
charge \(q = \pm 10 e\) makes
a potential \(V_\ell(0,-t)\) at \(z=-t\)
that is less than \(kT_b/|e|\),
this electric field is approximately
\beq
\bos{E} \approx - V_\ell(0,0) 
\, \frac{\bos{\hat z}}{t}
\label {E 2}
\eeq
which is larger than 
if the potential \(V_\ell(0,-t)\)
were more significant.
\par
Electroporation is the formation
of pores in membranes by an electric field.
Depending on the duration of the
field and the type of cell,
an electric potential difference
across a cell's plasma membrane 
in excess of about 200 mV will
create pores.  
\par
There are two main
components to the energy of a pore.
The first is the line energy 
\(2 \pi r \gamma \)
due to the linear tension \(\gamma\),
which is of the order of 
\(10^{-11}\) J/m\@.  The second
is the electrical energy
\( - 0.5 \Delta C \pi r^2 (\Delta V)^2 \)
in which \(\Delta V\) is the voltage
across the membrane and
\(
\Delta C = C_w - C_\ell 
\)
is the difference between the specific
capacity per unit area \(C_w = \ep_w \ep_0/t\)
of the water-filled pore
and that \(C_\ell = \ep_\ell \ep_0 /t \)
of the pore-free membrane of thickness \(t\)\@.
There also is a small term due to the
surface tension \(\Sigma\) of the
plasma membrane of the cell, but
this term usually is negligible
since \(\Sigma\) is of the order of
\(2.5 \times 10^{-6}\) J/m\(^2\)~\citep{Dai1997}\@.
The energy of the pore in a plasma membrane is 
then~\citep{Chernomordik1979,Chernomordik1987,Chernomordik1988,Weaver1996,Chernomordik2001}
\beq
E(r) = 2 \pi r \gamma - \pi r^2 \Sigma 
- \thalf \pi r^2  \Delta C \, (\Delta V)^2.
\label {energy of pore}
\eeq
This energy has a maximum of
\beq
E(r_c) = \frac{2\pi \gamma^2}
{\Delta C (\Delta V)^2 + 2\Sigma}
\approx \frac{2 \pi \gamma^2}
{\Delta C (\Delta V)^2}
\label {Emax}
\eeq
at the critical radius 
\beq
r_c = \frac{2\gamma}
{\Delta C \, (\Delta V)^2 + 2\Sigma}
\approx \frac{ 2 \gamma }
{\Delta C \, (\Delta V)^2}.
\label {r critical}
\eeq
The chance of a pore forming
rises steeply with the magnitude of the voltage
and falls with the radius of the pore.

\par
\begin{table}
\caption{\label{tab:1-cpp table of voltages}
The voltage differences 
\( \Delta V_{CPP} + \Delta V_{NaCl} \) (mV)
across the plasma membrane induced by an R\(^N\)
oligoarginine as an \(\alpha\)-helix,
a random coil, 
or a \(\beta\)-strand and by the ions
of 156 mM Na\(^+\) and Cl\(^-\)
reacting to it.
The resting transmembrane potential
\(\mbox{} -120 < \Delta V_{cell} < - 20\) mV 
is not included.}
\label {1-cpp table of voltages with counterions}
\begin{ruledtabular}
\begin{tabular}{||c|c|c|c||} 
\, \(N\) 
& R\(^N\) \(\alpha\)-helix 
& R\(^N\) random coil
& R\(^N\) \(\beta\)-strand 
\\ \hline 
5 &  \(\mbox{}-144 \pm 4\)     & \(\mbox{}-154 \pm 3\)   & \(\mbox{}-148 \pm 4\) \\ \hline
6 &  \(\mbox{}-174 \pm 4\)   & \(\mbox{}-173 \pm 4\)  & \(\mbox{}-168 \pm 1\) \\ \hline
7 &  \(\mbox{}-206 \pm 4\)  & \(\mbox{}-201 \pm 3\)    & \(\mbox{}-189 \pm 2\) \\ \hline
8 &  \(\mbox{}-232 \pm 3\)  & \(\mbox{}-228 \pm 1\)    & \(\mbox{}-199 \pm 5\) \\ \hline
9 &  \(\mbox{}-256 \pm 6\) & \(\mbox{}-246 \pm 2\)    & \(\mbox{}-205 \pm 4\) \\ \hline
10 & \(\mbox{}-281 \pm 2\)  & \(\mbox{}-260 \pm 5\)   & \(\mbox{}-210 \pm 4\) \\ \hline
11 & \(\mbox{}-306 \pm 5\)   & \(\mbox{}-260 \pm 3\)   & \(\mbox{}-218 \pm 4\) \\ \hline
12 & \(\mbox{}-323 \pm 4\) & \(\mbox{}-261 \pm 4\)    & \(\mbox{}-213 \pm 2\) \\ \hline
\end{tabular}
\end{ruledtabular}
\end{table}

\par
If the transmembrane potential \(\Delta V\) is turned off
before the radius of the pore reaches
\(r_c\), then the radius \(r\) 
of the pore usually shrinks quickly
(well within 1 ms~\citep{Chernomordik1987}) to a radius so small
as to virtually shut-down the conductivity
of the pore.  This rapid closure occurs
because in (\ref{energy of pore})
the energy \(2\pi r\gamma\) dominates
over \(- \pi r^2 \Sigma\), 
the surface tension
\(\Sigma\) being negligible.
Such a pore is said to be reversible.
But if \(\Delta V\) remains on when
\(r\) exceeds the critical
radius \(r_c\), then the pore usually
will grow and lyse the cell;
such a pore is said to be irreversible.
\par
The formula (\ref{r critical}) provides
an upper limit on the radius of 
a reversible pore.  This upper limit
drops with the square of the transmembrane
voltage \(\Delta V\) from \(r_c = 3.6\) nm for \(\Delta V = - 200\) mV,
to 1.6 nm for \(\Delta V = - 300\), and 
to 0.9 nm for \(\Delta V = - 400\) mV\@.
\par
The time \( t_\ell \) for a pore's radius
to reach the critical radius \(r_c\) is 
the time to lysis; it
varies greatly and apparently randomly even
within cells of a given kind. 
In erythrocytes,
its mean value drops by nearly
an order of magnitude with each
increase of 100 mV in the
transmembrane potential~\citep{Chernomordik1987}
and is about a fifth of a second
when \(\Delta V = - 300\) mV\@.
\par
The chance of a potential \(\Delta V\) 
forming a pore of radius \(r\)
is proportional to
the Boltzmann factor \(\exp(-E(r)/(kT))\)\@.
The higher the potential \(\Delta V\) and
the narrower the pore,
the greater the chance of pore formation.

\subsection{The Model
\label{The Model}}

In the model advanced in 
references~\citep{Cahill2009,Cahill2010},
a polyarginine sticks
to the cell membrane as its
guanidinium groups electrostatically
interact with the
phosphate groups of the outer leaflet of
the phospholipid bilayer.
If the positive charge of the 
polyarginine exceeds about \(8|e|\),
then it can raise the
transmembrane potential above the
threshold for electroporation,
some \(-200\) mV~\citep{Dai1997,Chernomordik1979,Chernomordik1987,Chernomordik1988,Weaver1996,Chernomordik2001}\@.
\par
The transmembrane potential \(\Delta V\) 
is the sum
of three terms
\beq
\Delta V = \Delta V_{cell} + 
\Delta V_{CPP} + \Delta V_{NaCl}
\label {dV3}
\eeq
the resting transmembrane
potential \(\Delta V_{cell}\) of the cell
in the absence of CPPs, the transmembrane
potential \(\Delta V_{CPP}\) due to an
oligoarginine or other CPP, and the transmembrane
potential \(\Delta V_{NaCl}\) due to
the counterions of the extracellular medium.
The resting transmembrane
potential \(\Delta V_{cell}\) of the cell
varies between about 20 mV 
to more than 70 mV, depending upon
the type of cell.  Ideally, it is
measured experimentally.
The transmembrane
potential \(\Delta V_{CPP}\) due to an
oligoarginine or to some other
positively charged CPP may be determined
from the formulas (\ref{Vellt}--\ref{Vct0})
for the potential of a charge outside
a membrane~\citep{Cahill2010}\@.
The transmembrane
potential \(\Delta V_{NaCl}\) due to
the counterions of the extracellular medium
requires a Monte Carlo simulation of the 
Na\(^+\), Cl\(^-\), and other ions
of the extracellular medium in the
electrostatic potential \(V_{cell} + V_{CPP}\)\@.
This simulation was performed
in~\citep{Cahill2010} with the aid of
equations (\ref{Vellt}--\ref{Vct0})\@.
The lipid bilayer insulates the
extracellular  counterions from the
potential due to the counterions
of the cytosol, however, so their potential
may be neglected in the simulation
of the extracellular  counterions.
\par
The values of \(\Delta V_{CPP} + \Delta V_{NaCl}\)
found in Monte Carlo
simulations~\citep{Cahill2010}
of the salt around 
an R\(^N\) oligoarginine
are listed in the table.
A resting transmembrane potential 
\(-120 < \Delta V_{cell} < -20\) mV
should be added to these values of
\(\Delta V_{CPP} + \Delta V_{NaCl}\)
to obtain the full transmembrane potential 
\(\Delta V\)\@.
For \(N > 8\), 
the transmembrane potential
\(\Delta V\) can exceed \(- 200\) mV
which is 
enough~\citep{Dai1997,Chernomordik1979,Chernomordik1987,Chernomordik1988,Weaver1996,Chernomordik2001} 
to cause electroporation
in common eukaryotic cells.
\par
In references \citep{Cahill2009}
and \citep{Cahill2010},
it was pointed out that one way
to test the model advanced 
in those papers would be
to look for
the formation of reversible
pores by detecting transient (ms) changes in the 
conductance of membranes exposed
to CPPs such as R\(^9\)\@.
Such experiments have now been done.
Using the planar-phospholipid-bilayer method,
Herce \tit{et al}.\ 
found that R\(^9\)
induced transient ionic currents 
through model phospholipid 
membranes~\citep{Herce2009}\@.
They estimated that the mean radius
of these pores to be \(0.66\) nm,
which is safely below the limiting critical
radius (\ref{r critical}) of
between 0.9 and 3.6 nm
for the voltage range of 
\(-400\le \Delta V \le - 200\) mV\@.
Moreover, using the patch-clamp technique,
they found that R\(^9\)
induced transient ionic currents 
through the membranes of
both freshly isolated 
human umbilical-artery (HUA) 
smooth-muscle cells and 
cultured osteosarcoma cells~\citep{Herce2009}\@.
These experimental confirmations
of the predictions made in references
\citep{Cahill2009} and \citep{Cahill2010}
lend some support to the model
advanced in these papers and sketched
in this subsection, but they do not prove
that it is correct.
It is, in any case, a continuum model 
of a molecular effect, and it may  
be consistent with the one
simulated in reference~\citep{Herce2009}\@.

\section{Smart Drugs
\label{Smart Drugs}}

Oligoarginines and other cell-penetrating peptides
can carry cargos of up to
4000 Da across the lipid bilayer
of eukaryotic cells by molecular
electroporation~\citep{Cahill2009,Cahill2010}\@.
This ability to transduce cargos
that substantially exceed the nominal
restriction barrier of 500 Da
of the ``rule of five''~\citep{Lipinski1997}
makes possible therapies like those sketched
in section~\ref{Therapeutic Applications}\@.
\par
But we also need
a way of targeting cancer cells.
One way to do this has been suggested
by Tsien~\citep{Tsien2004}\@.
His idea is to attach a negatively
charged molecule, an anion, to the CPP-cargo
molecule by a polypeptide linker
that is cut primarily by peptidases
that are over-expressed by the
targeted cells.  The geometry he proposed is
\beq
\msf{anion\!-\!linker\!-\!CPP\!-\!cargo}
\label {one anion}
\eeq
in which the hyphens represent
covalent bonds.  The cargo is
the therapeutic compound.
It might
be one of the peptides of 
(\citep{Tsien2004}--\citep{Tuennemann2007}),
which consisted of between 8 and 33 amino acids,
or simply a cytotoxin. 
If the negative charge
of the anion lowers
total charge of the anion-linker-CPP-cargo
molecule below a few \(|e|\),
then it will not enter the cell
by transduction because, 
in the model~\citep{Cahill2009,Cahill2010}, 
its transmembrane potential 
\(\Delta V\) will be too small
to cause electroporation. 
(The anion-linker-CPP-cargo molecule
might enter the cell by endocytosis,
but then it would risk
destruction as the pH of the endosome
dropped to that of a lysosome.)
Metastatic cancer cells over-express
certain transmembrane peptidases.
So to target them, one would pick
a linker that is cut by 
one of these over-expressed
transmembrane peptidases.

\subsection{Better Geometries
\label{Better Geometries}}

One may improve the selectivity
of such a smart drug by attaching
two or more anions to the 
positively charged CPP-cargo
by two or more \tit{different} linkers
that are cut primarily by \tit{different} 
peptidases
that are over-expressed by the
targeted cells. 
A simple geometry for two anions and two
different linkers is
\beq
\msf{anion_1\!-\!linker_1\!-\!CPP\!-\!cargo\!-\!linker_2\!-\!anion_2}.
\label {two anions}
\eeq
\par
To attach more than two anions
by more than two different linkers,
one can prepare an oxidizing solution 
of two classes of molecules.
A simple example of a 
molecule of the first class is
the molecule (\ref{two anions})
with an extra cysteine (\tsf{C}) 
\beq
\msf{anion_1\!-\!linker_1\!-\!C\!-\!CPP\!-\!cargo\!-\!linker_2\!-\!anion_2}
\label {two anions and one C}
\eeq
where the hyphens represent 
covalent bonds.  
Other examples of first-class molecules
contain more cysteines to the left
and/or right of the \tsf{CPP-cargo} moiety.
\par
In a more compact notation
in which \(\tsf{A}\) stands for an anion
and \(\tsf{L}\) for a linker,
two more examples of first-class molecules are
\beq
\msf{A_1\!-\!L_1\!-\!C\!-\!C\!-\!CPP\!-\!cargo\!-\!C\!-\!C\!-\!L_2\!-\!A_2}
\label {two anions and 4 Cs}
\eeq
with four extra cysteines and
\beq
\msf{A_1\!-\!L_1\!-\!C\!-\!C\!-\!C\!-\!CPP\!-\!cargo\!-\!C\!-\!C\!-\!C\!-\!L_2\!-\!A_2}
\label {two anions and 6 Cs}
\eeq
with six.
First-class molecules must have their
cysteines closer than any of the linkers
to the CPP-cargo moiety;
otherwise the cutting of 
a single linker could set adrift
more than one anion.
\par
Molecules of the second class
consist of a cysteine covalently
fused to a linker that in turn
is covalently fused to an anion;
that is, second-class molecules
are of the form
\beq
\msf{C\!-\!L_k\!-\!A_k}
\label {second class}
\eeq
in which \(\msf{L_k}\) is a linker
cleavable by a peptidase over-expressed
by the target cells and \(\msf{A_k}\)
is an anion.
\par
\begin{figure}
\centering
\rotatebox{0}
{\includegraphics[width=3in]{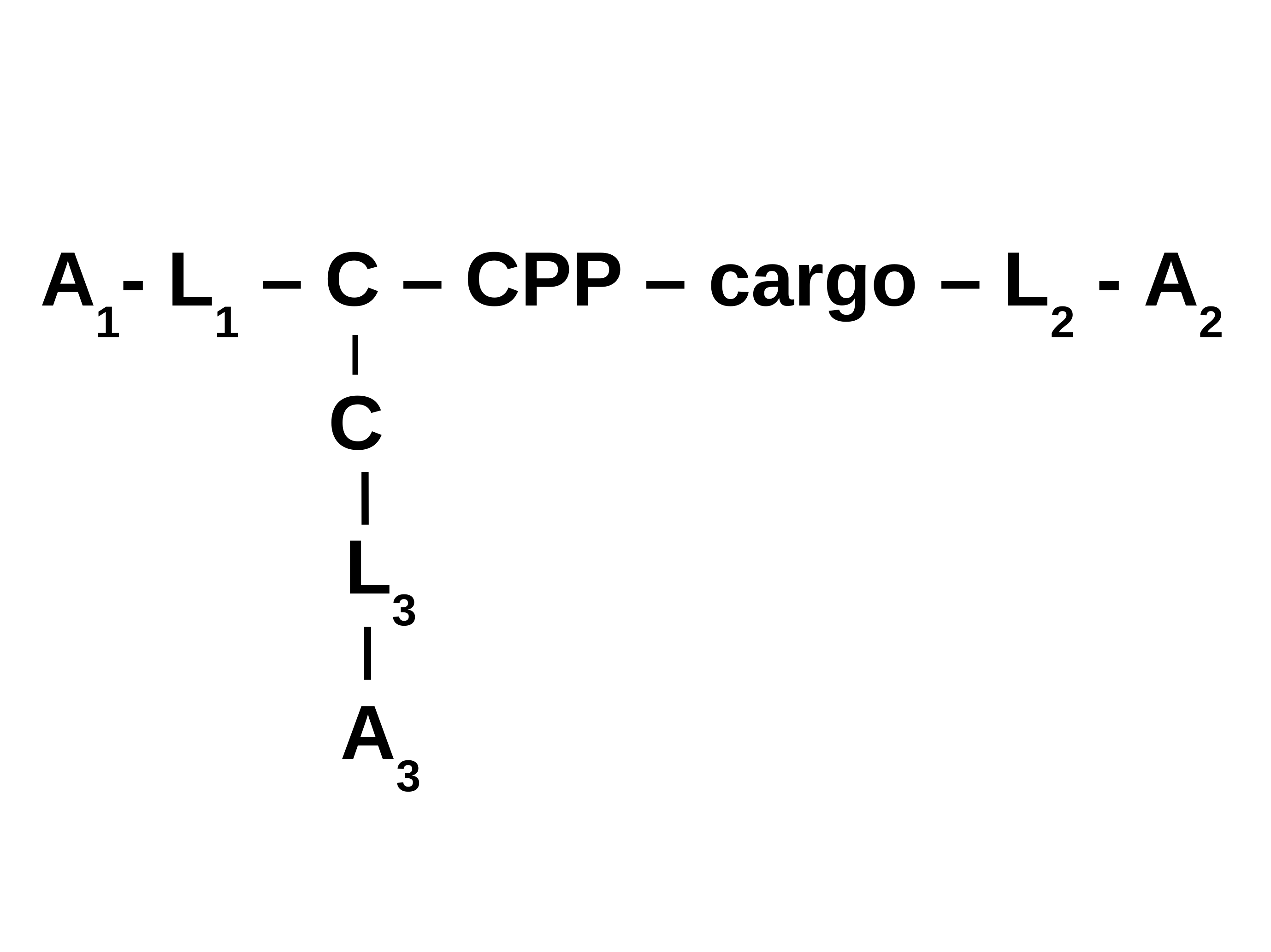}}
\caption{A smart drug consisting
of a CPP-cargo molecule and three
anions fused to it by three different
linkers that are cut primarily by
peptidases over-expressed mainly
by the targeted cells.
Here \tbf{\tsf{A}}, \tbf{\tsf{C}}, 
and \tbf{\tsf{L}} 
respectively stand for anion,
cysteine, and (8 aa) linker.
The hyphens and vertical lines
represent covalent bonds.}
\label {3c1Fig}
\end{figure}
\begin{figure}
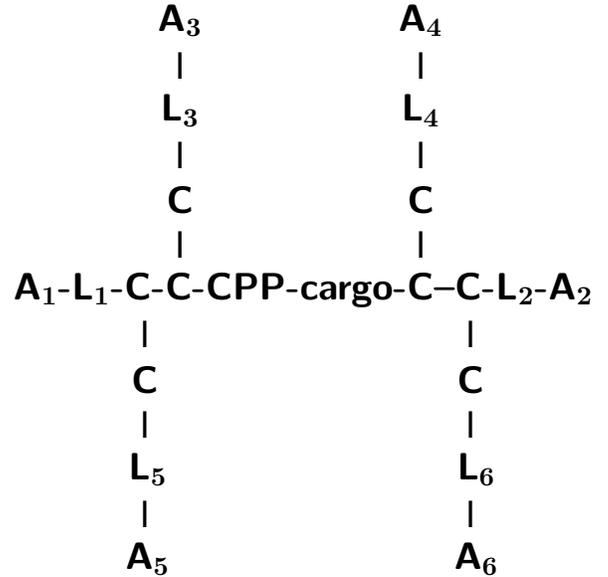

\centering
\input fig4
\caption{A smart drug consisting
of a CPP-cargo molecule and six
anions fused to it by six different
linkers that are cut primarily by
peptidases over-expressed mainly
by the targeted cells.
\tbf{\tsf{A}}, \tbf{\tsf{C}}, \tbf{\tsf{L}},
hyphens, and vertical lines are 
as in Fig.~\ref{3c1Fig}\@.}
\label {fig4Fig}
\end{figure}

In an oxidizing solution of these
two classes of molecules,
a cysteine of a second-class molecule
can form a disulfide bond with
a cysteine in a first-class molecule.
Thus, smart drugs 
can form in which three or more
anions are fused to the CPP-cargo
molecule by linkers that can be cut
by peptidases over-expressed by the
target cells.
For instance, an oxidizing solution
of the first-class molecule (\ref{two anions and one C})
and the second-class molecule (\ref{second class})
would make the smart drug shown in Fig.~\ref{3c1Fig}\@.
Similarly, an oxidizing solution
of the first-class molecule (\ref{two anions and 4 Cs})
and the second-class molecule (\ref{second class})
for \(k = 3\), 4, 5, and 6
would form the smart drug of Fig.~\ref{fig4Fig}\@.
\par
Of course, undesirable disulfide
bonds also would form.  One way
to suppress unwanted disulfide bonds
is to titrate
a dilute reducing solution
of different second-class 
molecules (\ref{second class}) into
an oxidizing solution of 
first-class molecules
such as (\ref{two anions and one C},
\ref{two anions and 4 Cs},
or \ref{two anions and 6 Cs})\@.
The order of the linkers
and anions does not matter
as long as they are different
and are cut primarily by
peptidases that are over-expressed
mainly by the target cells.
\par
To achieve maximum specificity,
it is desirable that the negative charge
of each anion reduce
the positive charge of the CPP
to \(4|e|\) or less, for in this
case, the CPP-cargo molecule
is not transduced until all the
linkers are cleaved.  Thus,
if the CPP is the oligoarginine
R\(^9\) with charge \(9|e|\),
then a smart drug like that
in Fig.~\ref{fig4Fig} would
have six anions of charge
\(-5|e|\) for a total charge 
of \(q = - 21 |e|\)\@.
Such smart drugs 
would be negatively charged.
\par
By using well-chosen linkers,
disulfide bonds,
CPP's, and cell-penetrating molecules (CPM's) 
such as cationic lipids, one can make
a very wide variety of smart drugs.

\subsection{Linkers
\label{Linkers}}

Linkers are essential
to the selectivity
of the smart drugs sketched 
in this section.
Each linker must be a short
peptide that is cut primarily
by a peptidase that is 
over-expressed mainly by
the targeted cells.
\par
MT1-MMP is a 
type-1 transmembrane proteinase
essential for
skeletal development, metastasis, and angiogenesis.
The Burnham Institute's cut database~\cite{Burnham}
lists 37 eight-aa-long sequences 
that have been shown by experiment
to be cut by MT1-MMP and that are not known
to be cut by other human peptidases:
\textsc{psqg-qkve, npmg-sepv, gyfg-dpla,
nlag-ilke, glrg-lqgp, lrrl-lglf,
aveg-sgks, dlsl-ispl, lisp-laqa,
geyr-tnpe, psqg-qkve, tkrd-lals,
rvlg-lete, rllg-lfge, dpfr-lqct,
lppg-lplt, ppsy-lgdr, qlyg-gesg,
nffp-rkpk, dpsa-imap, qglk-wqhn,
fciq-nytp, aepw-tvrn, qqly-gges,
pqpr-ttsr, aqlg-vmqg, mdet-mkel,
kayk-sele, llil-sdvn, lils-dvnd, 
dshs-lttn, lrgd-fssa, nmid-aatl,
kaiq-ltyn, glrg-lqgp, sser-ssts},
and \textsc{tsgg-yify}\@.
MT1-MMP cuts these 37 octapeptides
at the hyphens.
Although 37 sequences may seem 
so numerous as to be unselective,
in fact, they are 37 out
of \(20^8 = 25,600,000,000\) possible
8-aa sequences.
\par
The ADAMs are membrane proteins
that have both a disintegrin domain 
and a metalloprotease domain.
ADAM9, 10, 12, 15, and 17 
have been found in cancer cells.
ADAM17 (aka~\textsc{tace}) 
sheds and/or processes \textsc{tnf}\(\alpha\),
\textsc{tgf}\(\alpha\), \textsc{app}, amphiregulin, 
p75\textsc{tnfr}, p55\textsc{tnfr}, 
\textsc{trance}, 
L-selectin, IL-6 receptor,
IL-1 receptor II, Notch1 receptor, 
growth hormone-binding protein, \textsc{muc}1, and 
transmembrane collagen XVII~\cite{Duffy2003}\@.
The Burnham cut database~\cite{Burnham}
lists nine eight-amino-acid-long sequences 
that have been shown by experiment
to be cut by \textsc{adam17}
and that are not known
to be cut by any other human peptidase:
\textsc{dlla-vvaa, nsar-segp, kldk-sfsm,
lpvq-dsss, wtgh-stlp, rlrr-glaa,
ksmk-thsm, rveq-vvkp}, and \textsc{vaaa-vvsh}\@.
The cuts occur at the hyphens.
\par
Prostate-specific membrane antigen (PSMA)
is a glutamate carboxypeptidase II 
highly expressed by  
prostate epithelial cells
and by the neovasculature of many tumor types
but not by endothelial cells in 
normal tissue~\cite{Denmeade2004}\@.
Denmeade \textit{et al.}\  have found~\cite{Denmeade2004}
that the substrate 
\beq
\msf{APA\!-\!D\!-\!E}^*\msf{E}^*\msf{D\!-\!E}
\label {APA}
\eeq
is stable in blood but is
cut---at the second hyphen---
by PSMA\@.
Here APA is 4-\textsc{n}[\textsc{n}-2,4diamino-6-pteridinyl-methyl)-\textsc{n}-methyl\-amino-benzoate], and
the asterisks represent
\(\gamma\)-linkages, while
the hyphens stand for
the usual \(\alpha\)-linkages.

\subsection{Cell-Penetrating Molecules
\label{Cell-Penetrating Molecules}}
\begin{figure}
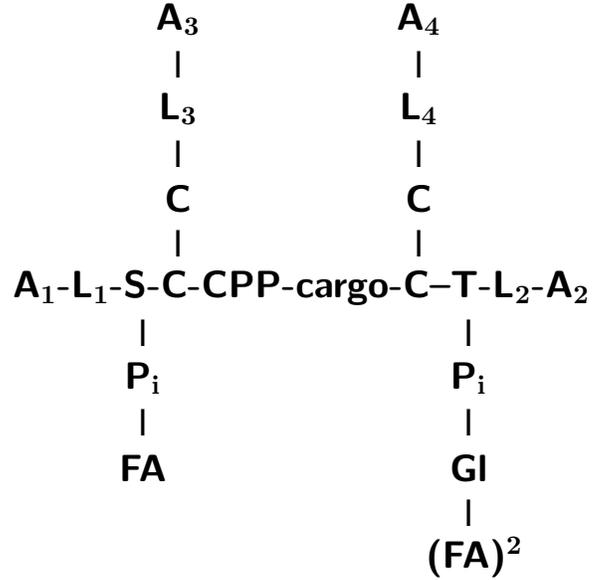

\centering
\input fig5
\caption{A smart drug consisting
of a CPP-cargo molecule and four
anions fused to it by four different
linkers that are cut primarily by
peptidases over-expressed mainly
by the targeted cells.
\tsf{\tbf{P\(_\mbf{\msf{i}}\)}} stands for
inorganic phosphate,
\tbf{\tsf{FA}} for fatty acid, and
\tbf{\tsf{Gl}} for glycerol.
\tbf{\tsf{A}}, \tbf{\tsf{C}}, \tbf{\tsf{L}},
hyphens, and vertical lines are 
as in Fig.~\ref{3c1Fig}\@.}
\label {fig5Fig}
\end{figure}
Two features of oligoarginines
allow them to penetrate through
cell membranes:  a positive charge
exceeding \(8|e|\) and several
guanidinium groups.  Thus any molecule
with several guanidinium groups and
a positive charge exceeding \(8|e|\)
may be as good a 
cell-penetrating molecule (CPM)
as the oligoarginines considered
in this paper.  
\par
Cationic lipids can penetrate
cell membranes; they form
another category of cell-penetrating molecules.
One design for a cationic lipid has
one or several serines \tsf{S}
and/or threonines \tsf{T} to the left
and/or right of the \tsf{CPP-cargo} moiety.
Thus an example of a first-class molecule 
for a cationic-lipid CPM is
\beq
\msf{A_1\!-\!L_1\!-\!S\!-\!C\!-\!CPP\!-\!cargo\!-\!C\!-\!T\!-\!L_2\!-\!A_2}.
\label {cationic lipid}
\eeq
To complete this first-class molecule
one would use the same kind
(\ref{second class}) of 
second-class molecule and
third-class molecules
that are either a fatty acid \tsf{FA}
fused to an 
inorganic phosphate group \(\msf{P_i}\)
by an acyl-phosphate bond
\beq
\msf{FA}\!-\!\msf{P_i}
\label {FAP}
\eeq
or a phospholipid 
without its polar head group
\beq
(\msf{FA})^2\!=\!\msf{Gl}\!-\!\msf{P_i}
\label {PL'}
\eeq
in which \tsf{Gl} is a glycerol.
One would make the cationic-lipid CPM
by fusing the second-class molecules
to the cysteines \tsf{C} by disulfide bonds
and the third-class molecules to the
hydroxyl groups of 
serines \tsf{S} and threonines \tsf{T}
by ester bonds.
The resulting smart drug is
illustrated in Fig.~\ref{fig5Fig}\@.
Such cationic-lipid smart drugs
with various numbers of anions,
linkers, and lipids may
be able to carry therapeutic cargos
across the cell membrane.

\subsection{Stability
\label{Stability}}

Blood has peptidases that cut
peptides and RNAses that cut RNAs,
but there are some biochemical
tricks that can increase the stability
of peptides and RNAs in smart drugs.
One may use right-handed amino acids
to frustrate peptidases.
A peptide made of
right-handed amino acids
in reverse order can have similar
biochemical properties to its
normally ordered 
left-handed twin~\citep{Goodman1993}\@.
Such \tit{retro-inverso} peptides
are more stable in the body
and were used in several of the
therapeutic experiments sketched
in section~\ref{Therapeutic Applications}\@.
RNAs made with 2'-\tit{O}-methyl-modified 
nucleotides with phosphorothioate linkages
resist RNAses~\citep{Stoffel2005}\@.
The wide class of cell-penetrating molecules
may have members
that are stable in the human body
and unlikely
to stimulate an immune response.

\subsection{Ligands and Antibodies
\label{Ligands and Antibodies}}

What is really needed for specificity
is knowledge of what distinguishes the surface
of a cancer cell from that of 
a healthy cell.  Cancer cells
over-express many receptors, such as
the CXC chemokine receptor 4 (CXCR4),
the specific neurokinin-1 (NK-1) receptor,
folate receptors (FRs), the 
G-protein-coupled protease-activated receptor 
PAR-1, and 
receptors for many regulatory peptides.
Metastatic cancer cells over-express
many transmembrane peptidases.
A reasonably complete characterization
of the distinguishing features
of the surfaces of cancer cells
would advance the development
of compounds that bind to these
features.  Such an 
artificial feature-ligand (FL) 
could guide a CPP-cargo molecule
to cancer cells.  The simplest
geometry would be
\beq
\msf{FL\!-\!L_1\!-\!CPP\!-\!cargo\!-\!L_2\!-\!A}
\label{FL CPP}
\eeq
in which the linkers \(L_1\)
and \(L_2\) are cleavable
by peptidases overexpressed
by the cancer cell that 
overexpresses the feature \(F\)\@.

\section{Summary
\label{Summary}}

In a model, which recently has received some
experimental support,
polyarginines carry cargos
by molecular electroporation
across the cell membrane.
They and other
cell-penetrating peptides (CPPs)
as well as cell-penetrating molecules (CPMs),
such as cationic lipids,
can transduce therapeutic cargos 
that greatly exceed the 500 Da restriction
barrier.
They may solve part 
of the drug-delivery problem.
\par
The use of well chosen linkers
and anions can help target 
cancer cells and spare
healthy ones.  The development
of really smart drugs, however,
requires a more complete
characterization of the 
differences between the surfaces
of cancer cells and those of normal cells.

\begin{acknowledgments}

I am grateful to Leonid Chernomordik 
for tips about electroporation, to
Gisela T\"{u}nnemann for sharing her data, 
to Sergio Hassan for advice about the
NaCl potential,
to Pavel Jungwirth for advice on 
guanidinium groups,
to John Connor and Karlheinz Hilber for 
explaining the status of measurements
of the membrane potential of mouse myoblast cells,
to Paul Robbins for sending me some
of his images, and to Jean Vance for
information about mammalian cells
deficient in the synthesis of 
phosphatidylserine.
Thanks also to S.~Atlas, B.~Becker,
H.~Berg, S.~Bezrukov, H.~Bryant, 
P.~Cahill, D.~Cromer, E.~Evans, 
A.~E. Garcia, B.~Goldstein, G.~Herling,
T.~Hess, S.~Koch,
V.~Madhok, M.~Malik, A.~Parsegian, 
B.~B. Rivers, K.~Thickman, T.~Tolley, and J.~Thomas
for useful conversations, and to  
K.~Dill, S.~Dowdy, S.~Henry, K.~Hilber, 
A.~Pasquinelli, 
B.~Salzberg, D.~Sergatskov, L.~Sillerud, 
B.~Smith, A.~Strongin, R.~Tsien, J.~Vance,
and A.~Ziegler 
for helpful e-mail.
\end{acknowledgments}

\bibliography{bio,physics,chem}
\end{document}